\documentclass[journal=jacsat,manuscript=article]{achemso}
\usepackage[version=3]{mhchem} 
\usepackage{siunitx}
\usepackage{multirow}
\usepackage{svg} 

\newcommand{\mr}[2]{\multirow{#1}{*}{#2}}

\author{Corin C. Wagen}
\email{corin@rowansci.com}
\author{Jonathon E. Vandezande}
\affiliation[Rowan Scientific]
{Rowan Scientific, Boston, MA}

\title[vDZP Paper]{The vDZP Basis Set Is Effective For Many Density Functionals}

\abbreviations{DFT,BSSE,BSIE}

\begin{document}

\begin{abstract}
In recent years, ``composite" density-functional-theory-based methods comprising specially optimized combinations of functionals, basis sets, and empirical corrections have become widely used owing to their robustness and computational efficiency, but the bespoke nature of these methods makes them challenging to develop. Here, we report that the recently reported vDZP basis set can be used in combination with a wide variety of density functionals to produce efficient and accurate results comparable to those obtained with composite methods, but without any method- or correction-specific reparameterization. This result enables rapid quantum chemical calculations to be run with a variety of density functionals without the typical errors incurred by small basis sets.
\end{abstract}

\section{Introduction}
The utility of quantum chemistry is limited by the inevitable tradeoff between the runtime of a calculation and the accuracy of the results obtained. In many domains, the applicability of quantum chemical calculations is determined not by the intrinsic error of the simulation but by the speed at which these calculations can be conducted and the size of the systems addressable. Accordingly, developing new methods that balance speed and accuracy in a Pareto-efficient manner is a crucial challenge facing computational chemists today.

Molecular quantum chemical calculations typically describe electron density through the linear combination of atom-centered Gaussian basis functions, and the choice of this basis set is key to both the speed and accuracy of the resulting calculation.\cite{nagy_basis_sets} The size of a basis set is often described in terms of $\zeta$ (zeta; the symbol traditionally used to denote basis-function exponents): single-$\zeta$ ``minimal" basis sets contain only a single basis function per atomic orbital, double-$\zeta$ basis sets contain two basis functions per atomic orbital, and so forth. Most basis sets today employ ``contracted" Gaussians, in which a single basis function is actually a linear combination of individual ``primitive" Gaussian functions designed to more closely mimic the shape of the true hydrogenic wavefunctions.\cite{nagy_basis_sets}

Small basis sets typically suffer from various pathologies: the electron density can be poorly described (basis-set incompleteness error, or BSIE) and interaction energies are often overestimated as fragments ``borrow" adjacent basis functions from each other (basis-set superposition error, or BSSE)\cite{Huzinaga1985}. These errors have been shown to cause dramatically incorrect predictions of thermochemistry, geometries, and barrier heights.\cite{Papajak2011, Boese2003, Kruse2012} Accordingly, conventional wisdom holds that triple-$\zeta$ basis sets or larger are required for accurate energy calculations, as demonstrated in this quote from a recent guide to best practices in computational chemistry:\cite{Bursch2022}

\begin{quote}
Therefore, DZ [double-$\zeta$] basis sets (like 6-31G** or def2-SVP) are no longer sufficient [for high-quality results], and we strongly advise against using them, except if they are part of purpose-made composite schemes. However, even in combination with full counterpoise corrections$\dots$ the residual BSSE and BSIE of DZ basis sets can be substantial. Thus, we generally recommend at least TZ [triple-$\zeta$] basis sets, which often yield results reasonably close to the basis set limit.
\end{quote}

Unfortunately, increasing the number of basis functions dramatically increases the runtime of the resulting calculation. Triple-$\zeta$ basis sets are substantially slower than double-$\zeta$ basis sets—in a recent benchmark study conducted by Folmsbee and Hutchison,\cite{Folmsbee2020} increasing the basis set from double-$\zeta$ (def2-SVP) to triple-$\zeta$ (def2-TZVP) caused calculation runtimes to increase more than five-fold. As a result, many calculations run on large or conformationally flexible systems still employ double-$\zeta$ basis sets for practical reasons, despite the known loss in accuracy that results.

One resolution to this unfortunate dilemma is the development of ``composite" density-functional-theory (DFT) methods which use highly optimized combinations of functional, basis set, and empirical corrections to achieve significant speed increases relative to typical methods. Since 2013, Stefan Grimme and co-workers have developed a suite of these methods, which have seen widespread adoption by the computational chemistry community.\cite{Sure2013, pbeh3c, Brandenberg2014, b973c, r2scan3c, wb97x3c} While early composite methods featured numerous fine-tuned empirical corrections—including a short-range basis correction for electronegative elements, a geometric counterpoise correction to correct for BSSE, and reparameterization of the underlying functional—the latest composite DFT method, $\omega$B97X-3c, employs only the D4 dispersion correction and a specially developed double-$\zeta$ basis set.\cite{wb97x3c} This basis set, vDZP, extensively uses effective core potentials to remove core elections and relies on deeply contracted valence basis functions optimized on molecular systems to minimize BSSE almost down to the triple-$\zeta$ level.\cite{Chan2023}

We hypothesized that the benefits of the vDZP basis set might not be limited to the $\omega$B97X-3c method, but instead could allow for efficient and low-cost calculations with a variety of other density functionals. In this work, we investigate the general applicability of vDZP by investigating the combination of vDZP with four additional functionals: B3LYP, M06-2X, B97-D3BJ, and r$^2$SCAN. In every case, we find that vDZP can produce highly accurate methods without any reparameterization of the functional or additional corrections beyond the now-standard empirical dispersion correction. We then examine B97-D3BJ and r$^2$SCAN functionals in more depth, and show that vDZP-based methods have speed and accuracy similar to existing composite methods in a variety of different benchmarks, while substantially outperforming conventional double-$\zeta$ basis sets.

\section{Methodology}

All computations were conducted with Psi4 1.9.1.\cite{Psi4} The default settings in Psi4 were modified somewhat: a (99,590) integration grid with ``robust" pruning, the Stratmann–Scuseria–Frisch quadrature scheme was employed,\cite{Stratmann1996} and an integral tolerance of $10^{-14}$ was used throughout. Density fitting was employed for all calculations, and a level shift of 0.10 Hartree was applied to accelerate SCF convergence. For the ROT34 benchmark, geometry optimizations were run using geomeTRIC 1.0.2.\cite{Wang2016}

Due to the documented absence of fluorine in Psi4's internal implementation of vDZP, a custom basis-set file was used which adds the missing basis functions for fluorine.

Timing studies were run on a dedicated ``Premium CPU-Optimized'' Digital Ocean Droplet with 8 Intel Cascade Lake processors and 16 GB of memory, and Psi4 was given 12 GB of memory.

\section{Results and discussion}

To assess the general applicability of vDZP, we selected four commonly used functionals from the ``Charlotte’s Web" of possible combinations of exchange and gradient approximations, with dispersion corrections as applicable: B97-D3BJ (GGA), r$^2$SCAN-D4 (meta-GGA), B3LYP-D4 (hybrid GGA), and M06-2X (hybrid meta-GGA). (We also included $\omega$B97X-D4, the range-separated hybrid functional explored in the original $\omega$B97X-3c paper.) We evaluated all functionals in combination with vDZP on the expansive GMTKN55 main-group thermochemistry benchmark set, which is now standard for quantifying the accuracy of new DFT methods, and compared these results to reference values obtained with the large (aug)-def2-QZVP basis set.\cite{GMTKN55} (Due to certain documented errors in Psi4's effective-core-potential implementation, the subsets NBPRC, FH51, DC13, C60ISO, and HEAVY28 were omitted.)

\begin{table}[ht]
    \centering
    \setlength{\tabcolsep}{4.6pt}
    \begin{tabular}{l|c*6{|S[table-format=2.2]}}
				&				&	{Basic}	&					& {Barrier} &\multicolumn{2}{c|}{NCI}&			\\
Functional			&	{Basis Set}	& {Properties}	&{Isomerization}& {Heights}	& {Inter-}	& {Intra-}	& {WTMAD2}	\\ \hline \hline
\mr{2}{B97-D3BJ}	&	def2-QZVP	&		5.43	&		14.21	&	13.13	&	5.11	&	7.84	&	8.42 	\\
				&	vDZP		&		7.70    &		13.58	&	13.25	&	7.27	&	8.60	&	9.56 	\\ \hline
\mr{2}{r$^2$SCAN-D4}&	def2-QZVP	&		5.23	&		8.41	&	14.27	&	6.84	&	5.74	&	7.45 	\\
				&	vDZP		&		7.28	&		7.10	&	13.04	&	9.02	&	8.91	&	8.34 	\\ \hline
\mr{2}{B3LYP-D4}	&	def2-QZVP	&		4.39	&		10.06	&	9.07	&	5.19	&	6.18	&	6.42 	\\
				&	vDZP		&		6.20	&		9.26	&	9.09	&	7.88	&	8.21	&	7.87 	\\ \hline
\mr{2}{M06-2X}		&	def2-QZVP	&		2.61	&		6.18	&	4.97	&	4.44	&	11.10	&	5.68 	\\
				&	vDZP		&		4.45	&		7.88	&	4.68	&	8.45	&	10.53	&	7.13 	\\ \hline
\mr{2}{$\omega$B97X-D4$^a$}&def2-QZVP&		3.18	&		6.04	&	3.75	&	2.84	&	3.62	&	3.73 	\\
				&	vDZP		&		4.77	&		7.28	&	5.22	&	5.44	&	5.80	&	5.57 	\\ \hline
    \end{tabular}
    \caption{\centering  
Weighted errors for various properties in GMTKN55.\hspace{\textwidth}
$^a${$\omega$B97X-3x} utilizes a refit D4 correction.\cite{wb97x3c}
    }
    \label{tab:GMTKN55}
\end{table}

The results of this study are shown in Table \ref{tab:GMTKN55}. In every case, the overall accuracy of methods employing vDZP is only moderately worse than the accuracy of methods using the much larger (aug)-def2-QZVP basis set, suggesting that vDZP is a generally applicable low-cost basis set. To assess whether vDZP was overly tailored for the $\omega$B97X-3c composite method, as with many components of previous composite methods, we compared the difference in accuracy between vDZP and (aug)-def2-QZVP for each functional under study. We found that the difference in overall accuracy was in fact largest for $\omega$B97X-D4, suggesting that vDZP is indeed well-suited outside of the $\omega$B97X-3c composite method.

Although the above results suggest that vDZP is effective for main-group thermochemistry, we wanted to more rigorously compare the performance of vDZP-based methods to popular composite methods with bespoke features. Accordingly, we selected B97-D3BJ/vDZP and r$^2$SCAN-D4/vDZP for detailed investigation, and compared the performance of these methods on GMTKN55 to the existing r$^2$SCAN-3c and B97-3c composite methods.\cite{b973c, r2scan3c} To benchmark vDZP against other double-$\zeta$ basis sets, we also evaluated three commonly employed double-$\zeta$ basis sets: 6-31G(d), def2-SVP, and pcseg-1.

\begin{table}[ht]
    \centering
    \begin{tabular}{l|c|S[table-format=1.2]|S[table-format=2.2]|S[table-format=2.2]|S[table-format=1.2]|S[table-format=2.2]|S[table-format=1.2]}
		&			&	{Basic}		&				& {Barrier} &\multicolumn{2}{c|}{NCI}&			\\
Basis Set	& {$\zeta$}	& {Properties}	& {Isomerization}& {Heights}& {Inter-}	& {Intra-}	&	{WTMAD2}\\ \hline \hline
def2-QZVP	&	4		&	5.43		&	14.21		& 	13.13 	& 	5.11	& 	7.84	&	8.42	\\ \hline
mTZVP$^a$ 		&	3		&	7.34		&	23.70		& 	13.14 	& 	10.25	& 	8.43	&	11.72	\\ \hline
\textbf{vDZP}&	2		&	7.70		&	13.58		& 	13.25 	& 	7.27	& 	8.60	&	9.56	\\ \hline
pcseg-1		&	2		&	9.71		&	15.18		& 	17.31 	& 	18.71	& 	19.78	&	15.58	\\ \hline
6-31G(d)	&	2		&	11.64		&	16.79		& 	18.14 	& 	22.43	& 	21.62	&	17.64	\\ \hline
def2-SVP	&	2		&	11.46		&	15.93		& 	18.68 	& 	26.79	& 	25.42	&	19.17	\\ \hline
    \end{tabular}
    \caption{In-depth evaluation of basis-set effects for B97-D3BJ.}
    \label{tab:GMTKN55_B97-D3BJ}
\vspace{0.15cm}
\hspace{0.65cm}\footnotesize{$^a$B97-3c}\raggedright
\end{table}

\begin{table}[ht]
    \centering
    \begin{tabular}{l|c*6{|S[table-format=2.2]}}
		&			&	{Basic}	&				& {Barrier} &\multicolumn{2}{c|}{NCI}&	 		\\
Basis Set	& {$\zeta$}	&{Properties}&{Isomerization}& {Heights}&	{Inter-}&	{Intra-}&{WTMAD2}	\\ \hline \hline
def2-QZVP	&	4		&	5.23	&		8.41	&	14.27	&	6.84	&	5.74	&	7.45	\\ \hline
mTZVPP$^a$	&	3		&	6.44	&		6.85	&	13.86	&	6.41	&	5.57	&	7.36	\\ \hline
\textbf{vDZP}&	2		&	7.28	&		7.10	&	13.04	&	9.02	&	8.91	&	8.34	\\ \hline
pcseg-1		&	2		&	9.09	&		8.47	&	17.20	&	18.47	&	20.89	&	14.44	\\ \hline
6-31G(d)	&	2		&	10.63	&		12.37	&	18.42	&	19.62	&	21.97	&	16.16	\\ \hline
def2-SVP	&	2		&	10.89	&		11.40	&	18.77	&	21.77	&	24.73	&	17.12	\\ \hline
    \end{tabular}
    \caption{In-depth evaluation of basis-set effects for r$^2$SCAN-D4.}
    \label{tab:GMTKN55_r2SCAN-D4}
\vspace{0.15cm}
\hspace{0.65cm}\footnotesize{$^a$r$^2$SCAN-3c}\raggedright
\end{table}

The results of this study are shown in Tables \ref{tab:GMTKN55_B97-D3BJ} and \ref{tab:GMTKN55_r2SCAN-D4}. In every case, vDZP far outperforms the other double-$\zeta$ basis sets. While the overall error of the other double-$\zeta$ basis sets is approximately twice that of the underlying functional, vDZP closely approaches the underlying error of the functional. In general, the vDZP-based methods have similar performance to the fine-tuned composite methods: while r$^2$SCAN-3c still outperforms r$^2$SCAN-D4/vDZP, B97-D3BJ/vDZP is markedly superior to B97-3c, especially for large systems and intramolecular non-covalent interactions.

Since GMTKN55 focuses exclusively on main-group elements, we next evaluated the accuracy of vDZP-based methods on transition metals by evaluating the revMOBH35 benchmark set,\cite{MOBH35, revMOBH35} which assesses the ability of computational methods to predict barrier heights in organometallic systems (Table \ref{tab:revMOBH35}). We found that vDZP-based methods had comparable accuracy to that of the congeneric composite methods, with slightly increased errors observed in both cases. 

\begin{table}[ht]
    \centering
    \begin{tabular}{l|S[table-format=2.2]}
Method						&	{MAE (kcal/mol)} \\ \hline \hline
GFN2-xTB					&		11.2			\\ \hline
\textbf{B97-D3BJ/vDZP}		&		4.2				\\ \hline
B97-3c						&		3.6				\\ \hline
\textbf{r$^2$SCAN-D4/vDZP}	&		2.9				\\ \hline
r$^2$SCAN-3c				&		2.8				\\ \hline
$\omega$B97X-3c				&		3.2				\\ \hline
$\omega$B97X-D4/def2-QZVPP	&		2.4				\\ \hline
$\omega$B97X-V/def2-QZVPP	&		2.3
    \end{tabular}
    \caption{revMOBH35 benchmark results.}
    \label{tab:revMOBH35}
\end{table}

To assess the accuracy of vDZP-based methods for geometry optimizations, we employed the ROT34 rotational constant benchmark set,\cite{ROT34} which compares the rotational constants of optimized structures to experimental values (Table \ref{tab:ROT34}). vDZP-based methods outperformed composite methods both in terms of mean deviation (MD) and mean absolute deviation (MAD), and r$^2$SCAN-D4/vDZP performed similarly to high-quality PBE0/def2-TZVP results.

\begin{table}[ht]
    \centering
    \begin{tabular}{l|S[table-format=3.2]*3{|S[table-format=2.2]}}
Method				&	{MD}& {MAD}	& {MAX}	&	{RMSD}	\\ \hline \hline
GFN2-xTB			&	-1.5&	2.9	&	24.8&	6.6		\\ \hline
\textbf{B97-D3BJ/vDZP}&	0.2	&	0.5	&	1.2	&	0.5		\\ \hline
B97-3c				&	0.4	&	0.5	&	1.7	&	0.6		\\ \hline
\textbf{r$^2$SCAN-D4/vDZP}&	-0.2&0.3&	0.7	&	0.3		\\ \hline
r$^2$SCAN-3c		&	0.8	&	0.8	&	1.5	&	0.4		\\ \hline
$\omega$B97X-3c		&	0.1	&	0.5	&	2.7	&	0.7		\\ \hline
PBE0-D3/def2-TZVP	& -0.2	&	0.2	&	0.8	&	0.3
    \end{tabular}
    \caption{ROT34 benchmark results. All values in \%.}
    \label{tab:ROT34}
\end{table}

We also investigated the accuracy of vDZP-based methods at computing torsional energy profiles for drug-like molecules, an important and well-studied task in computer-assisted drug design.\cite{Behara2024} We evaluated a variety of methods on the TorsionNet206 dataset,\cite{Xiao2024} which scores energies against high-level CCSD(T)/def2-TZVP benchmarks (Table \ref{tab:TorsionNet206}). vDZP-based methods gave mean absolute errors of 0.40–0.45 kcal/mol, comparable to composite methods and conventional hybrid functionals with triple-$\zeta$ basis sets. In contrast, a commonly used double-$\zeta$ DFT method (B3LYP-D3BJ/6-31G(d)) gave much worse performance, demonstrating the importance of high-quality basis sets even for ``easy" calculations like torsional scans.

\begin{table}[ht]
    \centering
    \begin{tabular}{l|S[table-format=1.2]}
Method					&	{MAE (kcal/mol)}	\\ \hline \hline
GFN2-xTB				&	0.78	\\ \hline
B3LYP-D3BJ/6-31G(d)		&	0.58	\\ \hline
\textbf{r$^2$SCAN-D4/vDZP}&	0.46		\\ \hline
r$^2$SCAN-3c			&	0.43	\\ \hline
\textbf{B97-D3BJ/vDZP}	&	0.42	\\ \hline
B3LYP-D3/def2-TZVP		&	0.40	\\ \hline
M06-2X/def2-TZVP		&	0.39	\\ \hline
B97-3c					&	0.36	\\ \hline
$\omega$B97M-D3BJ/def2-TZVPPD&	0.15
    \end{tabular}
    \caption{TorsionNet206 benchmark results, ranked by MAE.}
    \label{tab:TorsionNet206}
\end{table}

Since the utility of low-cost DFT methods arises not only from accuracy but also from computational efficiency, we also examined the relative speed of vDZP-based methods and composite DFT methods. Unlike energetic results, timing results are inherently hardware- and software-dependent and are thus less universal, but nevertheless useful trends can often still be divined. We compared the speed of B97-D3BJ/vDZP and r$^2$SCAN-D4/vDZP to the composite methods B97-3c, r$^2$SCAN-3c, and $\omega$B97X-3c by measuring timing on a series of \textit{n}-alkanes. 

\begin{figure}
    \centering
    \includegraphics[width=0.6\linewidth]{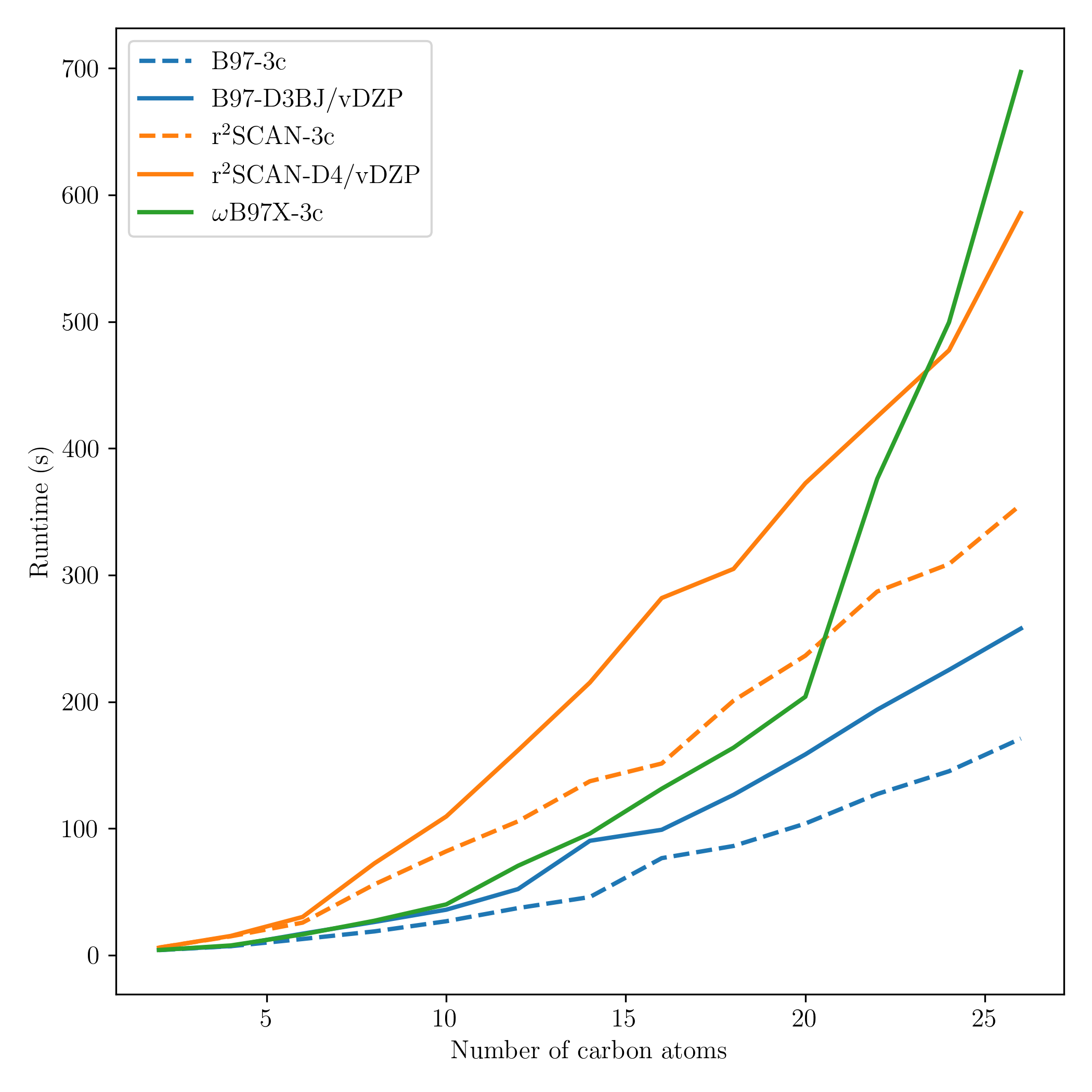}
    \caption{Timings for single-point energies of \textit{n}-alkanes.}
    \label{fig:timings}
\end{figure}

We found that the vDZP-based methods were on average 40\% slower than the corresponding composite methods (r$^2$SCAN-3c and B97-3c; Figure \ref{fig:timings}). This is somewhat surprising, given that vDZP contains fewer basis functions per atom than the triple-$\zeta$ mTZVP and mTZVPP basis sets used for B97-3c and r$^2$SCAN-3c, respectively. Two factors are likely responsible for this. First, vDZP extensively uses ECPs to describe core electrons, which simplifies computation of two-electron integrals but complicates computation of the one-electron Hamiltonian, and the overall effect on timing likely depends a great deal on the ECP implementation and nature of the system under study. Secondly, vDZP minimizes BSSE by using deeply contracted Gaussian functions, such that the highest angular momentum $L$ is small but the degree of contraction $K$ is high, so the total number of primitive basis functions remains large.\cite{wb97x3c} Existing two-electron-integral algorithms may be optimized for higher values of $L$ and lower values of $K$, and different algorithms like the Pople–Hehre axis-switch method may prove optimal for vDZP and other deeply contracted basis sets.\cite{Pople1978, Gill1994}

We also note that the extensive use of ECPs in vDZP can lead to substantial rate accelerations for systems with large numbers of heavy elements, since substantially fewer electrons will be modeled. For the particularly dramatic case of perbromo-\textit{n}-pentane, B97-3c is 3.4x slower than B97-D3BJ/vDZP, and r$^2$SCAN-3c is 2.7x slower than r$^2$SCAN-D4/vDZP. Overall, vDZP-based methods appear to have comparable efficiency to composite methods, and we anticipate that they can be made considerably faster if their use become commonplace.

\section{Conclusion}
Conventional wisdom in computational chemistry holds that existing basis sets are relatively optimal, and that that the tradeoff between speed and accuracy can only be resolved by tight coupling of methods, basis sets, and empirical corrections. This work suggests that this is false. Here, we demonstrate that the recently reported vDZP basis set is not limited to the specific $\omega$B97X-3c method for which it was originally reported, and instead can be combined with many different density functionals to produce fast and high-accuracy computational methods with Pareto efficiency comparable to bespoke composite methods.

More abstractly, the successes of vDZP detailed here suggest that there are considerable advances yet to come in basis-set optimization. The atypical features of vDZP—extensive use of large-core ECPs, deep contraction for valence orbitals, and parameter optimization on molecules, not atoms—are here demonstrated to create a robust and highly general solution for problems common to all electronic-structure-theory-based approaches. We anticipate that continued research into basis-set optimization will yield still faster and more accurate basis sets, allowing accurate quantum-chemical computations to scale to larger systems and timescales than ever before.

\begin{acknowledgement}
C.C.W. thanks Peter M. W. Gill and Todd Martínez for helpful discussions about quantum chemistry, many of which have indirectly percolated into this work.
\end{acknowledgement}

\begin{suppinfo}
The underlying data for the GMTKN55, ROT34, and revMOBH35 test sets, plus timing data on \textit{n}-alkanes and perbromo-\textit{n}-alkanes, is available in spreadsheet form.
\end{suppinfo}

\pagebreak
\bibliography{vDZP.bib}

\end{document}